# Co$_3$O$_4$ Nanocrystals on Graphene as a Synergistic Catalyst for Oxygen Reduction Reaction

Yongye Liang[§,1], Yanguang Li[§,1], Hailiang Wang[§,1], Jigang Zhou[2], Jian Wang[2], Tom Regier[2] and Hongjie Dai[*1]

[1]*Department of Chemistry, Stanford University, Stanford, CA 94305, USA and*

[2]*Canadian Light Source Inc., Saskatoon, SK, Canada.*

[§]These authors contributed equally to this work

*\* Correspondence should be addressed to* hdai@stanford.edu

**Catalysts for oxygen reduction and evolution reactions are at the heart of key renewable energy technologies including fuel cells and water splitting. Despite tremendous efforts, developing oxygen electrode catalysts with high activity at low costs remains a grand challenge. Here, we report a hybrid material of Co$_3$O$_4$ nanocrystals grown on reduced graphene oxide (GO) as a high-performance bi-functional catalyst for oxygen reduction reaction (ORR) and oxygen evolution reaction (OER). While Co$_3$O$_4$ or graphene oxide alone has little catalytic activity, their hybrid exhibits an unexpected, surprisingly high ORR activity that is further enhanced by nitrogen-doping of graphene. The Co$_3$O$_4$/N-doped graphene hybrid exhibits similar catalytic activity but superior stability to Pt in alkaline solutions. The same hybrid is also highly active for OER, making it a high performance non-precious metal based bi-catalyst for both ORR and OER. The**

**unusual catalytic activity arises from synergetic chemical coupling effects between $Co_3O_4$ and graphene.**

The increasing energy demand has stimulated intense research on alternative energy conversion and storage systems with high efficiency, low cost and environmental benignity[1-2]. Catalysts for oxygen reduction and evolution reactions are at the heart of key renewable energy technologies including fuel cells[3-4] and water splitting. Despite tremendous efforts, developing oxygen electrode catalysts with high activity at low costs remains a grand challenge. The current bottleneck of fuel cells reside in the sluggish ORR on the cathode side[4]. Pt or its alloys are the best known ORR catalysts. Due to the high cost of Pt and declining activity, alternative catalysts based on non-precious metals[5-7] and metal-free materials[8-10] have been actively pursued. On the other hand, the reverse reaction of ORR, OER or water oxidation plays an important role in energy storage such as solar fuel synthesis[2,11-12]. Ruthenium and iridium oxides in acidic condition and first row spinel and perovskite metal oxides in basic condition have been used to catalyze OER with moderate over-potentials (about 300-400 mV)[13].

It is highly challenging but desirable to develop efficient bi-functional catalysts for both ORR and OER, particularly for unitized regenerative fuel cell (URFC), a promising energy storage system that works as a fuel cell and in reversal as a water electrolyzer to produce $H_2$ and $O_2$ for feeding to the fuel cell[14-15].

Here, we show that $Co_3O_4$ nanoparticles, a material with little ORR activity by itself, when grown on reduced mildly-oxidized graphene oxide (rmGO) exhibits surprisingly high performance in both ORR and OER in alkaline solutions. The hybrid exhibits comparable ORR catalytic activity to commercial carbon-supported Pt catalyst (20 wt% Pt on Vulcan XC-72, Pt/C) and superior stability. This leads to a new bi-functional catalyst for ORR and OER.

$Co_3O_4$/graphene hybrid was synthesized in solution[16-20] by a general two step method. In the first step, $Co_3O_4$ nanoparticles were grown on mildly oxidized GO sheets (mGO) freely suspended in solution by hydrolysis and oxidation of cobalt acetate ($Co(OAc)_2$) at $80^\circ C$ (see Supplementary Information (SI) for details of the synthesis). Controlled nucleation of $Co_3O_4$ on mGO sheets was achieved by reducing the hydrolysis rate of $Co(OAc)_2$ through adjusting ethanol/$H_2O$ ratio and reaction temperature. Subsequent hydrothermal reaction at $150^\circ C$ led to crystallization of $Co_3O_4$ and reduction of mGO (signaled by a shift in the optical absorbance peak of GO[21-22]) to form the $Co_3O_4$/rmGO hybrid. We also added $NH_4OH$ in our synthesis steps to mediate hydrolysis of $Co^{2+}$ and its oxidation[23], and obtained an N-doped hybrid material denoted as $Co_3O_4$/N-rmGO (see SI). The amount of $Co_3O_4$ in our hybrid materials was ~70 wt% (~20 at% of Co) revealed by thermal-gravimetric measurements.

Growth of $Co_3O_4$ nanocrystals on rmGO sheets was confirmed by scanning electron microscopy (SEM) for both $Co_3O_4$/N-rmGO (Fig. 1a) and $Co_3O_4$/rmGO (Fig. S1a). Transmission electron microscopy (TEM) revealed smaller particles in $Co_3O_4$/N-rmGO (~4-8 nm in size, Fig. 1b) than $Co_3O_4$/rmGO (~12-25 nm in size, Fig. S1b), attributed to $NH_3$ coordination with cobalt cations in reducing $Co_3O_4$ particle size[23-24] and enhanced particle nucleation on N-doped rmGO (N-rmGO). High resolution TEM showed crystalline spinel structure of $Co_3O_4$ nanocrystals (Fig. 1c & Fig. S1c), consistent with X-ray diffraction (XRD) data (Fig. 1d & Fig. S1d). X-ray photoelectron spectroscopy (XPS) revealed 4 at% nitrogen in $Co_3O_4$/N-rmGO (Fig. 1e), but not in $Co_3O_4$/rmGO sample made without $NH_4OH$ (Fig. S2a). In a control experiment, we verified that N-dopants were on reduced GO sheets (not in $Co_3O_4$ nanocrystals, see Fig. S2b) due to hydrothermal reactions between functional groups on mGO and $NH_4OH$ in the solution[25]. High resolution XPS spectra of the N peak revealed pyridinic and pyrrolic nitrogen species in $Co_3O_4$/N-rmGO (Fig. 1e inset) and in N-rmGO (Fig. S2c).

To assess their ORR catalytic activity, our materials were first loaded (with the same mass loading) onto glassy carbon electrodes for cyclic voltammetry (CV) in $O_2$- vs. Ar-saturated 0.1 M KOH (see SI for experimental details). Free $Co_3O_4$ nanocrystals (size ~ 4-8 nm, similar to those grown on GO) or rmGO alone exhibited very poor ORR activity (Fig. S3). Remarkably, the $Co_3O_4$/rmGO hybrid showed a much more positive ORR onset potential (~0.88 V relative to the reversible hydrogen electrode (RHE), see SI for RHE calibration) and higher cathodic currents (Fig. 2a vs. Fig. S3), suggesting synergistic ORR catalytic activity of $Co_3O_4$ and rmGO in the hybrid.

We used rotating-disk electrode (RDE) measurements to reveal the ORR kinetics of our $Co_3O_4$/rmGO hybrid in 0.1 M KOH (Fig. 2b). The linearity of the Koutecky-Levich plots and near parallelism of the fitting lines suggests first-order reaction kinetics toward the concentration of dissolved oxygen and similar electron transfer numbers for ORR at different potentials[26] (Fig. 2b inset). The electron transfer number ($n$) was calculated to be ~3.9 at 0.60-0.75 V from the slopes of Koutecky-Levich plots[27] (Fig. 2b inset and see SI), suggesting $Co_3O_4$/rmGO hybrid favors a 4$e$ oxygen reduction process, similar to ORR catalyzed by a high quality commercial Pt/C catalyst measured in the same 0.1M KOH electrolyte ($n$ ~ 4.0 for Pt/C, see Fig. S4a, b).

The performance of our hybrid catalyst was greatly enhanced with $Co_3O_4$/N-rmGO made by adding $NH_4OH$ during synthesis to afford N-doping in rmGO. The CV curve of $Co_3O_4$/N-rmGO hybrid exhibited a more positive ORR peak potential and higher peak current (Fig. 2a) than $Co_3O_4$/rmGO. RDE measurement revealed an electron transfer number of ~4.0 at 0.60-0.75 V (Fig. 2c). The half wave potential at 1600 rpm was 0.83 V (Fig. 2c), similar to that of Pt/C (0.86 V) (Fig. S4a) and more positive than that of $Co_3O_4$/rmGO (0.79 V) (Fig. 2b). Importantly, N-doped graphene (N-rmGO) alone without $Co_3O_4$ exhibited low ORR activity (Fig. S3) with an electron transfer number of ~2.7 at 0.50 - 0.65 V, suggesting a dominant 2e reduction process (Fig. S5). Excellent ORR activity of the $Co_3O_4$/N-rmGO hybrid catalyst was also gleaned from the much smaller Tafel slope of 42 mV/decade at low over-potentials (Fig. 2d) than that measured with $Co_3O_4$/rmGO hybrid (54 mV/decade) in 0.1 M KOH.

To verify the ORR catalytic pathways of the hybrid materials, we performed rotating ring-disk electrode (RRDE) measurements to monitor the formation of peroxide species ($HO_2^-$) during the ORR process[28] (Fig. 3a). The measured $HO_2^-$ yields are below ~12% and ~6% for $Co_3O_4$/rmGO and $Co_3O_4$/N-rmGO respectively over the potential range of 0.45 to 0.80 V, giving electron transfer number of ~3.9 (Fig. 3b). This is consistent with the result obtained from the Koutecky-Levich plots based on RDE measurements, suggesting the ORR catalyzed by our hybrids is mainly 4*e* reduction.

We loaded our catalyst materials onto Teflon-treated carbon fiber paper (CFP) (at ~0.24 mg/cm$^2$ for all samples) to measure their steady-state iR-compensated polarization curves (Fig. 4). The Teflon treated porous CFP is highly hydrophobic, providing three phase contact point for oxygen, electrolyte and catalyst[29] (known as the gas diffusion layer) useful in actual fuel cells to minimize the gas diffusion limitation[30-31]. In 0.1 M KOH at 0.7 V vs. RHE (a typical half-cell cathodic potential in an operating fuel cell), our $Co_3O_4$/rmGO and $Co_3O_4$/N-rmGO hybrids afforded an ORR current density of ~12.3 mA/cm$^2$ and ~52.6 mA/cm$^2$ respectively (Fig. 4a), approaching that of the Pt/C catalyst (~68.0 mA/cm$^2$). The oxygen reduction currents of our hybrid catalysts were 1-3 orders of magnitude higher than $Co_3O_4$ (0.012 mA/cm$^2$), rmGO (0.19 mA/cm$^2$) or N-rmGO alone (3.5 mA/cm$^2$) (Fig. S6), further suggesting synergetic coupling effects between two catalytically non-active components in our hybrid for ORR catalysis.

In 1 M and 6 M KOH electrolytes, our $Co_3O_4$/N-rmGO ORR catalyst matched the performance of freshly loaded Pt/C catalyst in current density (Fig. 4b, c), accompanied by a positive shift in the ORR onset potential from 0.1 M KOH. The Tafel slope of kinetic current was down to ~37 mV/decade for $Co_3O_4$/N-rmGO in 1 M KOH (Fig. S7), amongst the smallest Tafel slopes afforded by ORR catalysts. Importantly, our hybrid exhibits superior durability than Pt/C catalyst in 0.1 M-6 M KOH with little decay in ORR activity over 10,000-25,000 s of continuous operation (Fig. 4d-f). In contrast, the Pt/C catalyst exhibited 20%-48% decrease in activity in 0.1-6 M KOH (Fig. 4d-f), giving lower long term ORR currents than the stable currents sustained by the $Co_3O_4$/N-rmGO hybrid catalyst.

We further compared the $Co_3O_4$/N-rmGO hybrid catalyst with other catalysts (Fig. S8). Although the Pt/C − 50% (50 wt% Pt on Vulcan XC-72), a state-of-the-art Pt catalyst, showed slightly higher activity (Fig. S8a), it suffered a significant 30% decrease in current density over 10,000 s of continuous operation in 1 M KOH (Fig. S8b). A commercial Pd/C catalyst (10% Pd on activated carbon) showed lower activity than $Co_3O_4$/N-rmGO hybrid and obvious decay in activity over time (by ~20% in 10,000 s). Non-precious metal-N/C catalysts have shown excellent ORR activity in acid[6] and in base[32-33]. To compare, we prepared a high quality Fe-N/C catalyst following Ref.[6] & [30] (see Fig.S9 for RDE data). Notwithstanding a more positive onset potential, the Fe-N/C catalyst showed slightly lower activity than our $Co_3O_4$/N-rmGO hybrid at around 0.7 V. Further, the Fe-N/C catalyst exhibited ~10% decreases in current density over 10,000 s continuous operation, suggesting lower stability/durability than the $Co_3O_4$/N-rmGO hybrid. Activity instability of metal-N/C catalysts in acidic electrolyte is also well known and has been a limiting factor to the practical use of this material.[6-7] Pt catalyst is known to gradually degrade over time due to surface oxides and particle dissolution and aggregation, especially in alkaline electrolyte used for alkaline fuel cells (AFCs)[4,34]. Long term stabilities of other ORR catalysts such as Ag in alkaline solutions are improved over Pt, but still suffer from deactivation and are below the targets for applications[29]. Since the lack of catalyst durability has been one of the major challenges for AFC, the excellent stability of our $Co_3O_4$/N-rmGO hybrid makes it promising for ORR and other important catalytic reactions in alkaline solutions[35].

We performed x-ray absorption near edge structure (XANES) measurements to glean the interactions between $Co_3O_4$ and GO in our hybrids (Fig. 1f). Compared to N-rmGO, $Co_3O_4$/N-rmGO hybrid showed a clear increase of carbon K-edge peak intensity at ~288 eV, corresponding to carbon atoms in graphene attached to oxygen or other species[36-37]. This suggested possible formation of interfacial Co-O-C and Co-N-C bonds in the $Co_3O_4$/N-rmGO hybrid. In the oxygen K-edge XANES, an obvious decrease in unoccupied O 2p–Co 3d hybridized state (~532 eV)[38] was observed (Fig. 1f inset) accompanied by an increase in the Co L-edge XANES (mapping of the unoccupied Co 3d projected state, see Fig. S10) peak[38] in the hybrids compared to pure $Co_3O_4$ nanocrystal, suggesting higher electron density at O site and lower electron density at Co site, consequently a higher ionic Co-O bonding in the hybrid[39]. Bond formation between $Co_3O_4$ and N-rmGO (as suggested by C K-edge XANES) and changes in the chemical bonding environment for C, O and Co atoms in the hybrid material are likely responsible for the synergistic ORR catalytic activity.

N-doping of GO could afford stronger coupling between Co and graphene in $Co_3O_4$/N-rmGO than in $Co_3O_4$/rmGO. N-groups on reduced GO serve as favorable nucleation and anchor sites for $Co_3O_4$ nanocrystals due to coordination with Co cations[24]. This is consistent with smaller $Co_3O_4$ nanocrystal size (higher degree of nucleation) on N-rmGO than on rmGO. Electronic effects of N-doping of graphene could also play a role to ORR. Note that Metal-N species is believed to be ORR-active sites in Fe- or Co-N/C catalysts[7] prepared at much higher temperatures (600-1000°C) with much lower metal loadings (< 1-2 at% of metal)[7]. Our hybrid catalysts exhibit higher stability and differ in low temperature solution-phase synthesis and that Co in our hybrid is in the form of oxides (see XPS in Fig. S2d) with a high Co loading of ~20 at%. Lowering Co loading to 3-10 at% in our $Co_3O_4$/N-rmGO catalyst led to systematic reduction in ORR activity (Fig. S11), suggesting that the active reaction sites in our hybrid materials could be Co oxide species at the interface with graphene.

The mechanism of ORR with our hybrid catalyst remains unclear. The Tafel slope of kinetic current down to ~37 mV/decade for $Co_3O_4$/N-rmGO in 1M KOH is among the lowest for spinel oxide ORR catalysts and close to 2.303(2RT/3F) V/decade (R, universal gas constant; F, faraday constant), suggesting protonation of $O_2^-$ on the active sites of catalyst as the rate limiting step[40]. A dual-site mechanism has been proposed for cobalt-polypyrrole/C ORR catalyst, in which oxygen is reduced to peroxide at Co-N-C sites and further reduced to $OH^-$ at $Co_xO_y$/Co sites[41]. A similar mechanism may be at work in our hybrid system. However, we note that $Co_3O_4$/rmGO without any N species has very similar ORR onset potential as $Co_3O_4$/N-rmGO, suggesting that the active sites may not directly involve N species in the hybrids but are enhanced by N doping of mGO. It is also found that physical mixtures of $Co_3O_4$ with rmGO or N-rmGO afforded much lower ORR activities than the corresponding hybrid material (Figure S12), suggesting synergistic coupling between $Co_3O_4$ and graphene is indispensible to the high ORR activity of the hybrid.

We also used the same method to prepare $Co_3O_4$ hybrids with other forms of carbon and compared their ORR performances (Fig. S13). The result indicated that high conductivity, high surface area, and suitable functional groups on carbon materials are important to the high activity of hybrid materials.

Lastly, we extended the potential of our hybrid electrode to 1.70 V vs. RHE to the water oxidation regime and evaluated electrocatalytic oxygen evolution reaction (OER). In 0.1 M KOH, the same sample used for ORR containing $Co_3O_4$/N-rmGO loaded on CFP (at $0.24mg/cm^2$) afforded higher OER currents than either free $Co_3O_4$ nanocrystals or Pt/C (Fig. 5a). Further, we loaded our $Co_3O_4$/N-rmGO catalyst onto Ni foam at ~1 $mg/cm^2$ and evaluated OER performance in 1 M KOH at room temperature (Fig. 5b). Our catalyst afforded a current density of 10 $mA/cm^2$ at a small overpotential of ~0.31 V (Fig. 5b) and a small Tafel slope down to 67 mV/decade (Fig. 4c), comparable to the performance of the best reported $Co_3O_4$ nanoparticle OER catalyst at the same loading[42]. N-doping of graphene did not affect OER activity as $Co_3O_4$/rmGO hybrid showed only slightly lower OER activity than $Co_3O_4$/N-rmGO (Fig. 5b). Stability test on glassy carbon electrode showed that both $Co_3O_4$/N-rmGO and $Co_3O_4$/rmGO hybrids are inherently stable during OER cycling (Fig. S14). These results make our hybrid material a powerful bi-functional catalyst for both oxygen reduction and water oxidation. Previously, manganese oxide was shown to be a bi-functional catalyst for ORR and OER[15]. Our $Co_3O_4$/N-rmGO catalyst outperforms manganese oxide with smaller over-potentials for both ORR and OER, presenting the highest performance non-precious metal based bi-functional catalyst.

In summary, while $Co_3O_4$ or graphene oxide alone has little catalytic activity for ORR, their hybrid materials exhibit unexpected, surprisingly high ORR activities in alkaline solutions comparable to fresh commercial Pt/C catalyst but far exceeding Pt/C in stability and durability. This presents a highly promising catalyst for alkaline fuel cells for which there has been a recent resurgence in interest with solutions to electrolyte carbonation[35]. We also demonstrated $Co_3O_4$/graphene hybrid as one of the rare and highest performance bi-functional catalyst for ORR and water oxidation/OER. Thus, synergistic coupling of nanomaterials opens up a brand new approach to advanced catalysts for energy conversion.

**Methods**

**Materials synthesis.** GO was made by a modified Hummers method, in which a 6 times lower concentration of $KMnO_4$ was used than that for Hummers' GO. In the first step of $Co_3O_4$/rmGO hybrid synthesis, $Co(OAc)_2$ aqueous solution was added into GO/ethanol dispersion at room temperature. The reaction was kept stirring for 10 h at $80^o$C. In the second step, the reaction mixture from first step was transferred to an autoclave for hydrothermal reaction at $150\,^o$C for 3 h. To synthesize $Co_3O_4$/N-rmGO, $NH_4OH$ was added after $Co(OAc)_2$ addition in the first step. See Supplementary Information for detailed experimental procedures.

**Electrochemical measurements.** Saturated calomel electrode (SCE) was used as the reference electrode in all measurements and it was calibrated with respect to reversible hydrogen electrode (see SI for details). Cyclic voltammetry was conducted in a three-electrode electrochemical cell using a graphite rod as the counter electrode. 12 μg of sample was loaded on the glassy carbon working electrode (3 mm in diameter). In the rotating disk electrode and rotating ring-disk electrode measurements, the working electrode was prepared by loading 20 μg of sample on a glassy carbon electrode of 5 mm in diameter (0.1 mg/cm$^2$). For the measurement on carbon fiber paper, the working electrode was prepared by loading 0.24 mg of sample on 1 cm$^2$ carbon fiber paper from its ethanol dispersion with Nafion (10% to sample). All the data from carbon fiber paper were iR-compensated. The scan rate was 5 mV/s for all electrochemical measurements.

**Characterizations.** Scanning electron microscopy (SEM) was carried out on an FEI XL30 Sirion scanning electron microscope. Transmission electron microscopy (TEM) was carried out on an FEI Tecnai G2 F20 transmission electron microscope. X-ray diffraction (XRD) was carried out on a PANalytical X'Pert instrument. XPS was carried out on an SSI S-Probe XPS Spectrometer. The XANES at the C K-edge, O K-edge and Co L-edge were obtained on the spherical grating monochromator (SGM) beamline (E/ΔE~5000) at the Canadian Light Source (CLS), a 2.9 GeV third generation synchrotron source.

**Acknowledgement**

We thank Dr. T. F. Jaramillo for insightful discussions. This work was supported in part by ONR. CLS is supported by the NSERC, NRC, CIHR of Canada, and the University of Saskatchewan.

**Author contributions:**

Y.L., Y.L., H.W. and H.D. conceived the project and designed the experiments. Y.L., Y.L. and H.W. performed the experiments. J.Z. J.W. and T. R. performed the XANES measurement and analysis. Y.L., Y.L. and H.L. analyzed the data. Y.L., Y.L. and H.D. co-wrote the paper. All authors discussed the results and commented on the manuscript.

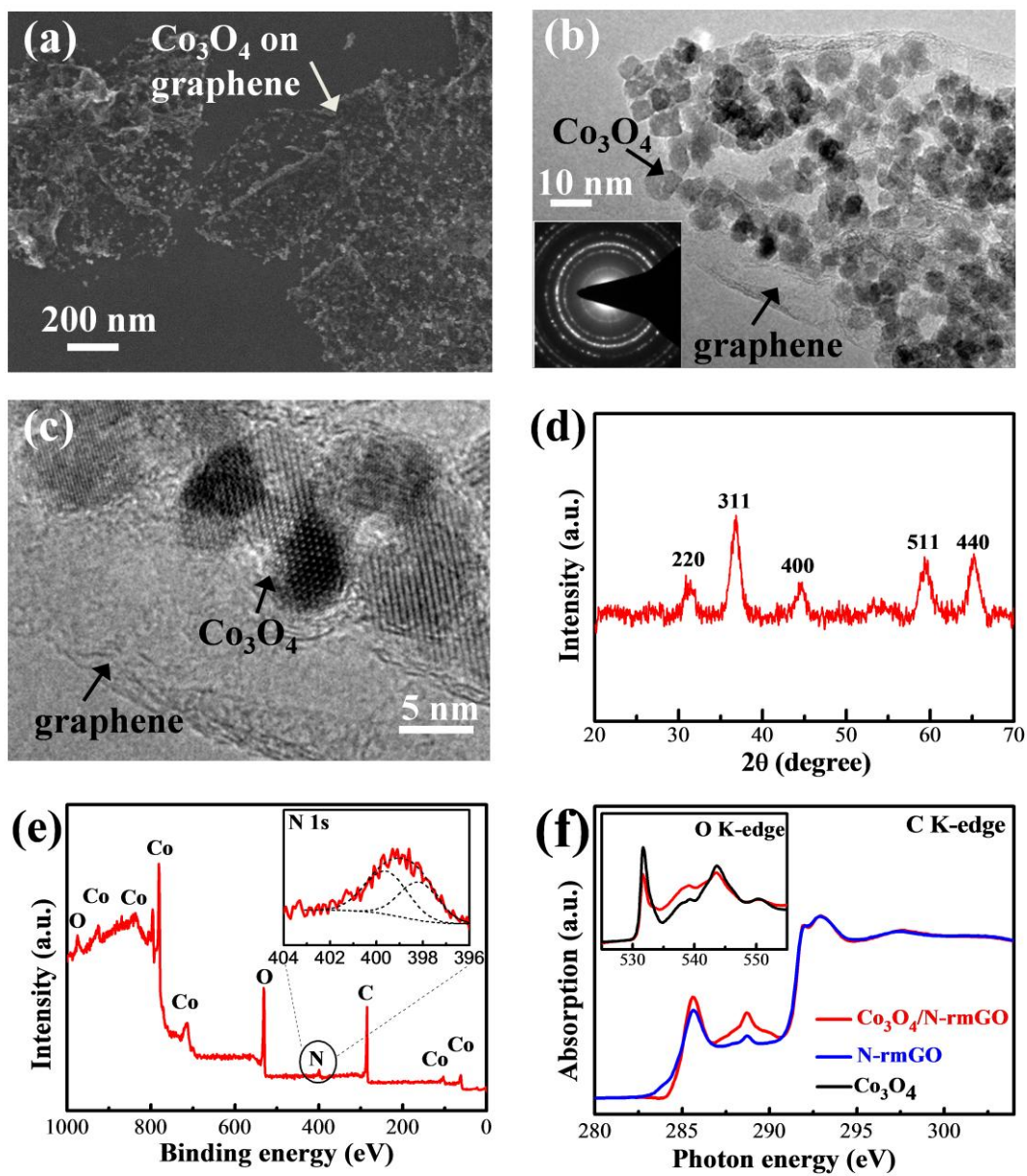

**Fig. 1**. Co₃O₄/graphene hybrid materials. (a) A SEM image of Co$_3$O$_4$/N-rmGO hybrid deposited on silicon substrate from a suspension in solution. (b) Low magnification and (c) high magnification TEM images of Co$_3$O$_4$/N-rmGO hybrid. (d) An XRD spectrum of a compacted film of Co$_3$O$_4$/N-rmGO hybrid. (e) An XPS spectrum of Co$_3$O$_4$/N-rmGO hybrid. Inset shows a high resolution N1s spectrum with the peak de-convoluted into pyridinic and pyrollic N peaks. (f) C K-edge XANES of N-rmGO (blue curve) and Co$_3$O$_4$/N-rmGO hybrid (red curve). Inset shows O K-edge XANES of Co$_3$O$_4$ (black curve) and Co$_3$O$_4$/N-rmGO hybrid (red curve).

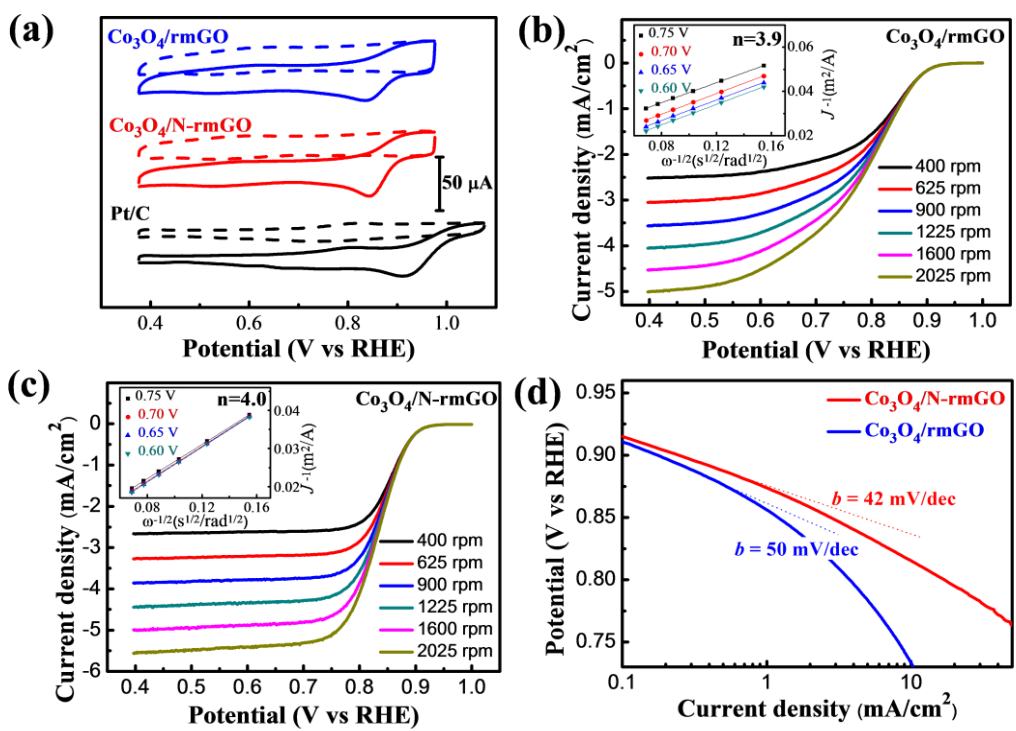

**Fig. 2**. Co₃O₄/graphene hybrid as oxygen reduction catalysts. (a) CV curves of Co₃O₄/rmGO hybrid, Co₃O₄/N-rmGO hybrid and Pt/C on glassy carbon electrodes in O₂-saturated (solid line) or Ar-saturated 0.1 M KOH (dash line). Catalyst loading was 0.17 mg/cm² for all samples. (b) Rotating-disk voltammograms of Co₃O₄/rmGO hybrid (loading~0.1 mg/cm²) and (c) Co₃O₄/N-rmGO hybrid (loading~0.1 mg/cm²) in O₂-saturated 0.1 M KOH with a sweep rate of 5 mV/s at different rotation rates indicated. The insets in (b) and (c) show corresponding Koutecky–Levich plots ($J^{-1}$ vs. $\omega^{-0.5}$) at different potentials. (d) Tafel plots of Co₃O₄/rmGO and Co₃O₄/N-rmGO hybrids derived by the mass-transport correction of corresponding RDE data (see Supplementary Information).

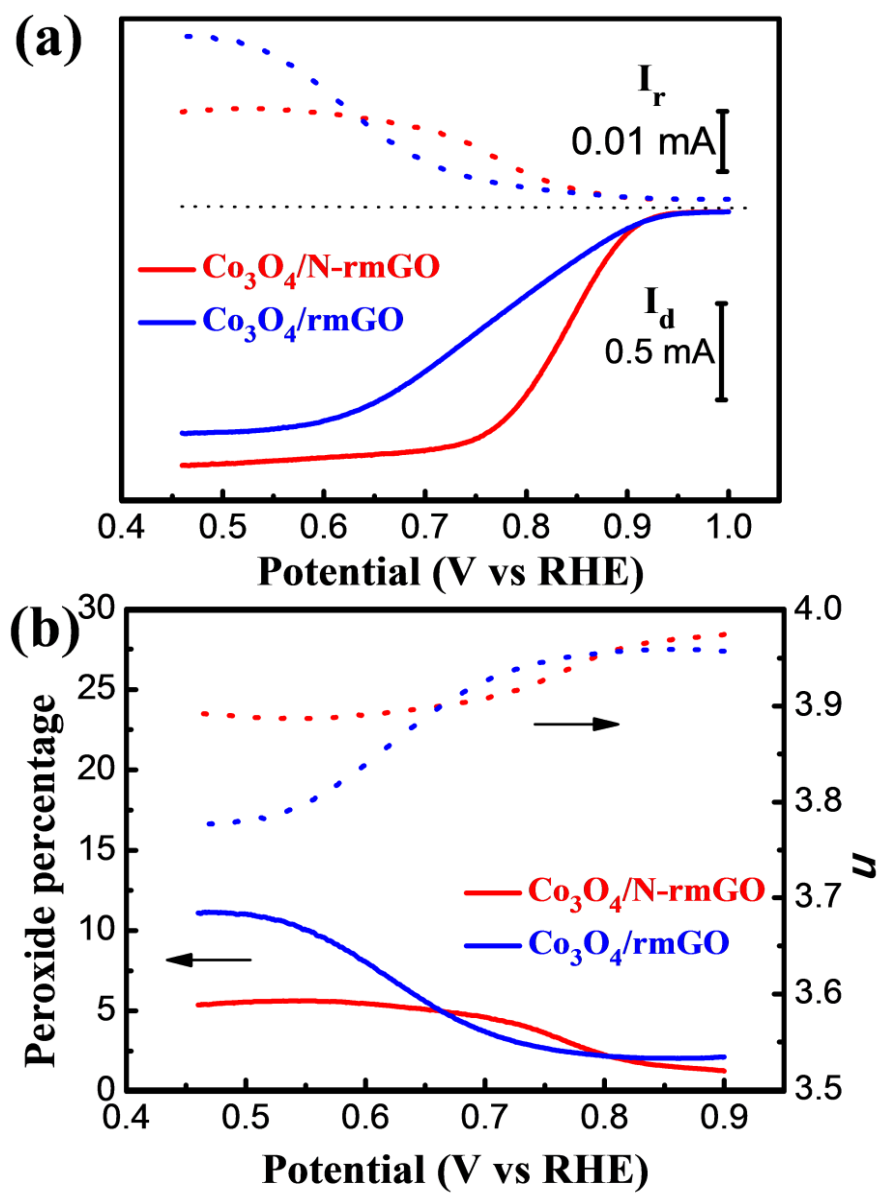

**Fig. 3**. Assessment of peroxide percentage in ORR catalyzed by hybrid catalysts. (a) Rotating ring-disk electrode voltammograms recorded with $Co_3O_4$/rmGO hybrid (loading~0.1 mg/cm$^2$) and $Co_3O_4$/N-rmGO hybrid (loading~0.1 mg/cm$^2$) in $O_2$-saturated 0.1 M KOH at 1600 rpm. Disk current ($I_d$) (solid line) is shown on the lower half and ring current ($I_r$) (dotted line) is shown on the upper half of the graph. The disk potential was scanned at 5 mV/s and the ring potential was constant at 1.5 V vs RHE. (b) Percentage of peroxide (solid line) and the electron transfer number ($n$) (dotted line) of $Co_3O_4$/rmGO and $Co_3O_4$/N-rmGO hybrids at various potentials based on the corresponding RRDE data in (a).

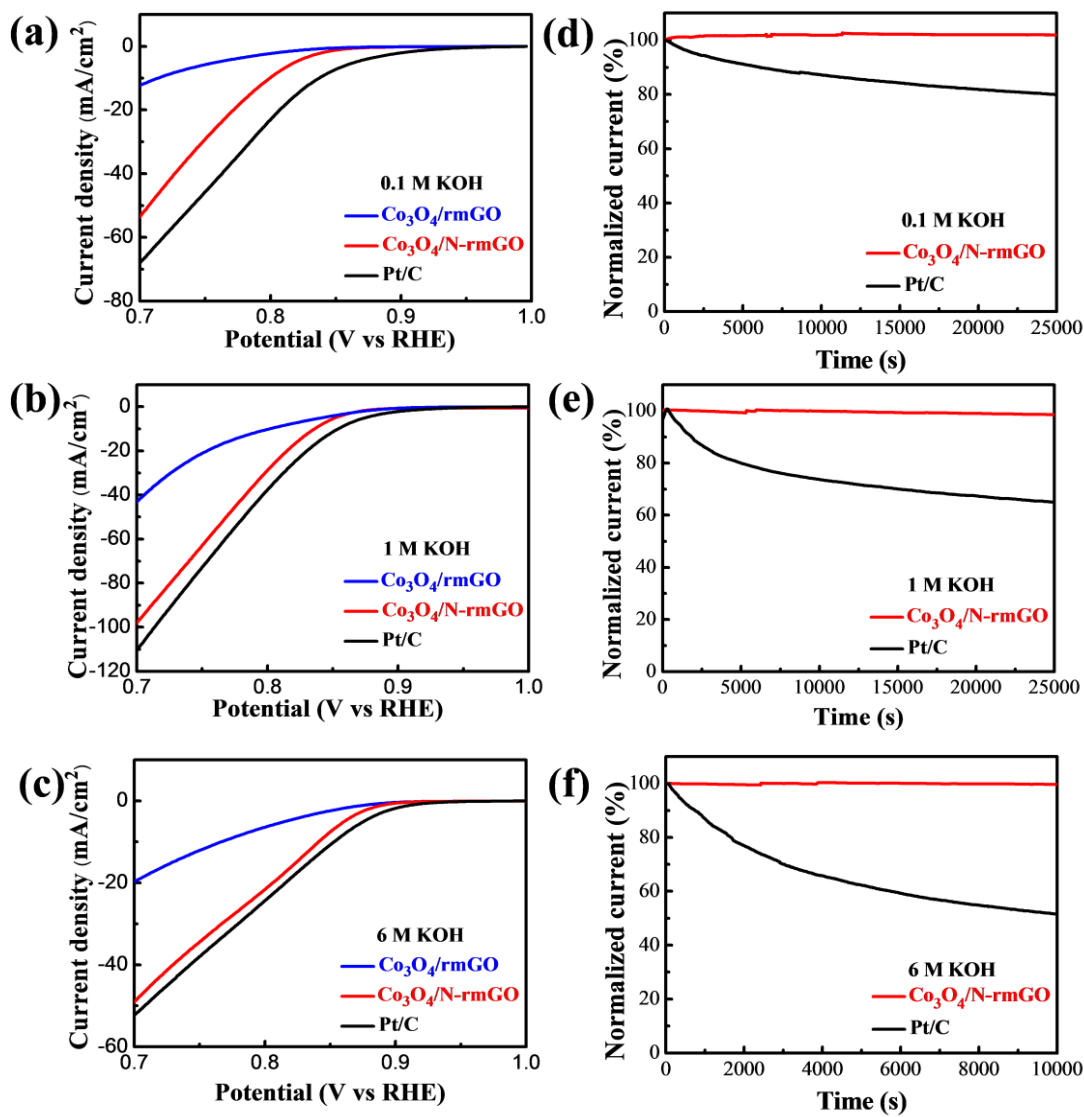

**Fig. 4.** ORR performance and stability of catalysts. (a)-(c) Oxygen reduction polarization curves of Co$_3$O$_4$/rmGO, Co$_3$O$_4$/N-rmGO and a high quality commercial Pt/C catalysts (catalyst loading ~0.24 mg/cm$^2$ for all samples) dispersed on carbon fiber paper (CFP) in O$_2$-saturated (a) 0.1 M KOH, (b) 1 M KOH and (c) 6 M KOH electrolytes respectively. (d)-(f) Chronoamperometric responses (percentage of current retained vs. operation time) of Co$_3$O$_4$/N-rmGO hybrid and Pt/C on carbon fiber paper electrodes kept at 0.70 V vs. RHE in O$_2$-saturated (d) 0.1 M KOH, (e) 1M KOH and (f) 6 M KOH electrolytes respectively. Co$_3$O$_4$/N-rmGO hybrid showed comparable ORR catalytic activity to Pt/C and superior stability in alkaline solutions.

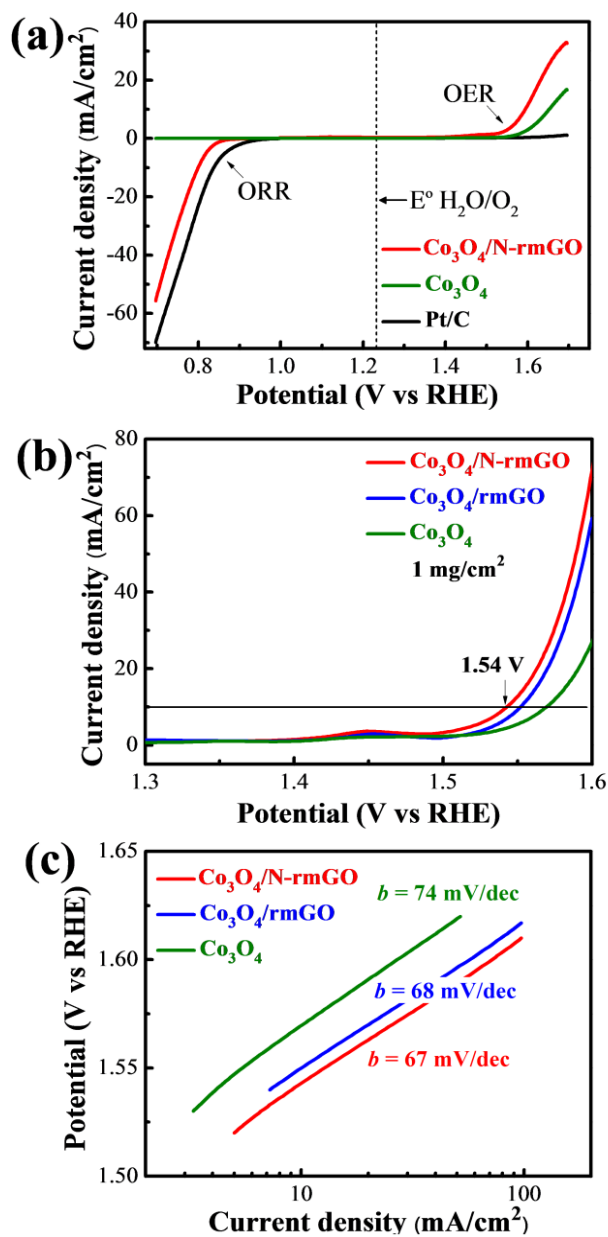

**Fig. 5**. Co₃O₄/graphene hybrid bi-functional catalyst for ORR and water oxidation (OER). (a) Oxygen electrode activities within the ORR and OER potential window of Co₃O₄/N-rmGO hybrid, Co₃O₄ nanocrystal and Pt/C catalysts (catalyst loading ~0.24 mg/cm² for all samples) dispersed on carbon fiber paper in O₂-saturated 0.1 M KOH. Co₃O₄/N-rmGO hybrid showed excellent catalytic activities for both ORR and OER. Free Co₃O₄ showed little ORR activity while Pt/C showed little OER activity. (b) Oxygen evolution currents of Co₃O₄/N-rmGO hybrid, Co₃O₄/rmGO hybrid and Co₃O₄ nanocrystal loaded onto Ni foam (to reach a high catalyst loading of ~1 mg/cm²) measured in 1 M KOH. (c) Tafel plots of OER currents in (b).

# Supplementary Information

## Co$_3$O$_4$ Nanocrystals on Graphene as a Synergistic Catalyst for Oxygen Reduction Reaction

### *Synthesis of Mildly Oxidized Graphene Oxide (mGO)*

mGO was made by a modified Hummers method using a lower concentration of oxidizing agent. Graphite flakes (1 g, Superior Graphite Co.) were grounded with NaCl (20 g) for 10-15 minutes. Afterwards, the NaCl was washed away by repeatedly rinsing with water in a vacuum filtration apparatus. The remaining graphite was dried in an oven at 70 ℃ for 30 minutes. The dried solid was transferred to a 250 ml round bottom flask. 23 ml of concentrated sulfuric acid was added and the mixture was stirred at room temperature for 24 hours. Next, the flask was heated in an oil bath at 40°C. 100 mg of NaNO$_3$ was added to the suspension and allowed to dissolve in 5 minutes. This step was followed by the slow addition of 500 mg of KMnO$_4$ (3 g for Hummers' GO), keeping the reaction temperature below 45°C. The solution was allowed to stir for 30 minutes. Afterwards, 3 ml of water was added to the flask, followed by another 3 ml after 5 minutes. After another 5 minutes, 40 ml of water was added. 15 minutes later, the flask was removed from the oil bath and 140 ml of water and 10 ml of 30% H$_2$O$_2$ were added to end the reaction. This suspension was stirred at room temperature for 5 minutes. It was then repeatedly centrifuged and washed with 5% HCl solution twice, followed by copious amounts of water. The final precipitate was dispersed in 100 ml of water and bath sonicated for 30 min. Any indispensable solid was crushed down by a centrifugation at 5000 rpm 5 minutes, and a brown homogeneous supernatant was collected.

### Synthesis of $Co_3O_4$/rmGO, $Co_3O_4$/N-rmGO Hybrids, rmGO, N-rmGO and free $Co_3O_4$ nanoparticles.

mGO was collected from the aqueous solution by centrifugation and redispersed in anhydrous ethanol (EtOH). The concentration of the final mGO EtOH suspension was ~0.33 mg/ml (concentration of our mGO stock suspension was determined by measuring the mass of the mGO lyophilized from a certain volume of the suspension). For the first step synthesis of hybrid without $NH_4OH$, 1.2 ml of 0.2 M $Co(OAc)_2$ aqueous solution was added to 24 ml of mGO EtOH suspension, followed by the addition of 1.2 ml of water at RT. The reaction was kept at 80 $^o$C with stirring for 10 h. After that, the reaction mixture from the first step was transferred to a 40 mL autoclave for hydrothermal reaction at 150$^o$C for 3 h. This hydrothermal step also reduced mGO to rmGO. The resulted product was collected by centrifugation and washed with ethanol and water. The resulting $Co_3O_4$/rmGO hybrid was ~20 mg after lyophilization.

To prepare $Co_3O_4$/N-rmGO hybrid with the addition of $NH_4OH$, the first step reaction mixture was prepared by adding 1.2 ml of 0.2 M $Co(OAc)_2$ aqueous solution to 24 ml of mGO EtOH suspension, followed by the addition of 0.50 ml of $NH_4OH$ (30% solution) and 0.70 ml of water at RT. The following steps were the same as above. The resulting $Co_3O_4$/N-rmGO hybrid was ~20 mg after lyophilization.

The mass ratio of graphene in the hybrid was determined by thermal-gravimetric analysis, in which the hybrid material was heated in air at 500℃ for 2 hours and a weight loss of ~30 % was measured. This corresponded to the removal of graphene from the hybrid by oxidation. $Co_3O_4$ was about 70% by mass (~20% by atom) in our hybrid.

rmGO was made through the same steps as making $Co_3O_4$/rmGO without adding any Co salt in the first step. N-rmGO was made through the same steps as making $Co_3O_4$/N-rmGO without adding any Co salt in the first step. This produced N-doped reduced GO with N clearly resolved in the GO sample by XPS (Fig.S2b). Free $Co_3O_4$ nanoparticle was made through the same steps as making $Co_3O_4$/N-rmGO without adding any mGO in the first step.

### *Sample preparation for SEM, TEM and XRD*

SEM samples were prepared by drop-drying the samples from their aqueous suspensions onto silicon substrates. TEM samples were prepared by drop-drying the samples from their diluted aqueous suspensions onto copper grids. XRD samples were prepared by drop-drying the samples from their aqueous suspensions onto glass substrates.

### *XANES Measurements*

XANES were recorded in the surface sensitive total electron yield (TEY) with use of specimen current. Data were first normalized to the incident photon flux $I_0$ measured with a refreshed gold mesh at SGM prior to the measurement. After background correction, the XANES are then normalized to the edge jump, the difference in absorption coefficient just below and at a flat region above the edge (300, 565 and 800 eV for C, O and Co respectively).

*Electrochemical Measurements*

**1. Cyclic voltammetry (CV).** 5 mg of catalyst and 16-106 µl (16 µl for hybrids or Pt/C, 15% of Nafion to catalyst ratio; 106 µl for N-rmGO or rmGO, 100% of Nafion to catalyst ratio) of 5 wt% Nafion solution were dispersed in 1 ml of 3:1 v/v water/isopropanol mixed solvent by at least 30 min sonication to form a homogeneous ink. Then 2.4 µl of the catalyst ink (containing 12 µg of catalyst) was loaded onto a glassy carbon electrode of 3 mm in diameter (loading ~ 0.17 mg/cm$^2$). Cyclic voltammetry (using the pontentiostat from CH Instruments) was conducted in a home-made electrochemical cell using saturated calomel electrode as the reference electrode, a graphite rod as the counter electrode and the sample modified glassy carbon electrode as the working electrode. Electrolyte was saturated with oxygen by bubbling $O_2$ prior to the start of each experiment. A flow of $O_2$ was maintained over the electrolyte during the recording of CVs in order to ensure its continued $O_2$ saturation. The working electrode was cycled at least 5 times before data were recorded at a scan rate of 5mVs$^{-1}$. In control experiments, CV measurements were also performed in Ar by switching to Ar flow through the electrochemical cell.

**2. Rotating disk electrode (RDE) measurement.** For the RDE measurements, catalyst inks were prepared by the same method as CV's. 4 µl ink (containing 20 mg catalyst) was loaded on a glassy carbon rotating disk electrode of 5 mm in diameter (Pine Instruments) giving a loading of 0.1 mg/cm$^2$.The working electrode was scanned cathodically at a rate of 5 mVs$^{-1}$ with varying rotating speed from 400 rpm to 2025 rpm. Koutecky–Levich plots ($J^1$ vs. $\omega^{-1/2}$) in the insets of Figure 2 of the main text were analyzed at various electrode potentials. The slopes of their best linear fit lines were used to calculate the number of electrons transferred ($n$) on the basis of the Koutecky-Levich equation[1]:

$$\frac{1}{J} = \frac{1}{J_L} + \frac{1}{J_K} = \frac{1}{B\omega^{1/2}} + \frac{1}{J_K}$$

$$B = 0.62nFC_o(D_o)^{2/3}\nu^{-1/6} \qquad\qquad J_K = nFkC_o$$

where $J$ is the measured current density, $J_K$ and $J_L$ are the kinetic- and diffusion-limiting current densities, $\omega$ is the angular velocity, $n$ is transferred electron number, $F$ is the Faraday constant, $C_o$ is the bulk concentration of $O_2$, $\nu$ is the kinematic viscosity of the electrolyte, and $k$ is the electron-transfer rate constant. For the Tafel plot, the kinetic current was calculated from the mass-transport correction of RDE by:

$$J_K = \frac{J \times J_L}{(J_L - J)}$$

**3. Rotating ring-disk electrode (RRDE) measurement.** For the RRDE measurements, catalyst inks and electrodes were prepared by the same method as RDE's. The ink was dried slowly in air and the drying condition was adjusted by trial and error until a uniform catalyst distribution across the electrode surface was obtained. The disk electrode was scanned cathodically at a rate of 5 mVs$^{-1}$ and the ring potential was constant at 1.5 V vs RHE. The % $HO_2^-$ and the electron transfer number (n) were determined by the followed equations[2]:

$$\% \; HO_2^- = 200 \times \frac{I_r/N}{I_d + I_r/N}$$

$$n = 4 \times \frac{I_d}{I_d + I_r/N}$$

where $I_d$ is disk current, $I_r$ is ring current and N is current collection efficiency (*N*) of the Pt ring. N was determined to be 0.40 from the reduction of $K_3Fe[CN]_6$.

**4. Oxygen electrode activities on carbon fiber paper.** For measurements on carbon fiber paper, the working electrode was prepared by loading ~0.24 mg of catalyst (for hybrid catalysts and Pt/C) on 1 $cm^2$ carbon fiber paper (purchased from Fuel Cell Store) from its 1 mg/ml ethanol dispersion with a 1:10 Nafion-to-catalyst ratio. It was cycled at least 20 times between 0 and -0.4 V vs SCE before data were recorded at a scan rate of $5mVs^{-1}$ for ORR measurement. To obtain both ORR and OER activities in 0.1 M KOH, the working electrode was scanned from -0.3 V to 0.7 V vs SCE after ORR measurement. Multiple cycles were recorded for each sample. The initial anodic sweep was showed in Figure 4a. All the data from carbon fiber paper were iR-compensated.

**5. RHE calibration** We used saturated calomel electrode (SCE) as the reference electrode in all measurements. It was calibrated with respect to reversible hydrogen electrode (RHE). The calibration was performed in the high purity hydrogen saturated electrolyte with a Pt wire as the working electrode. CVs were run at a scan rate of 1 mV $s^{-1}$, and the average of the two potentials at which the current crossed zero was taken to be the thermodynamic potential for the hydrogen electrode reactions.

a) 0.1 M KOH

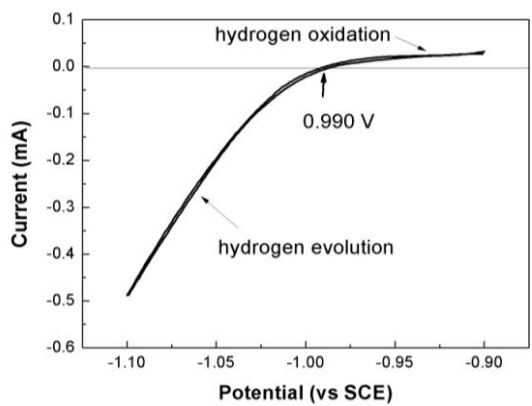

So in 0.1 M KOH, E (RHE) = E (SCE) + 0.990 V.

b) 1 M KOH

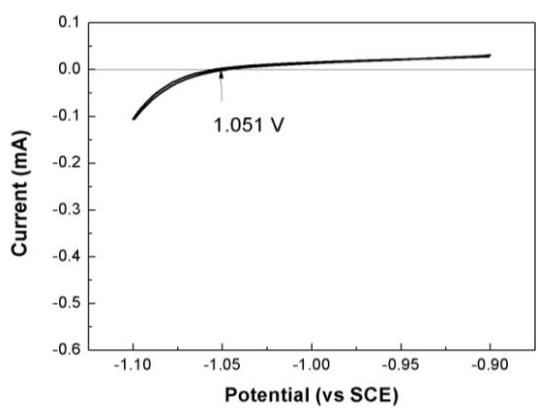

So the 1 M KOH, E (RHE) = E (SCE) + 1.051 V.

c) 6 M KOH

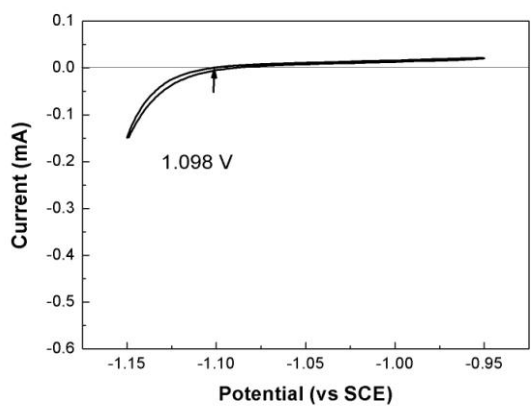

So in 6 M KOH, E (RHE) = E (SCE) + 1.098 V.

*Supplementary Figures*

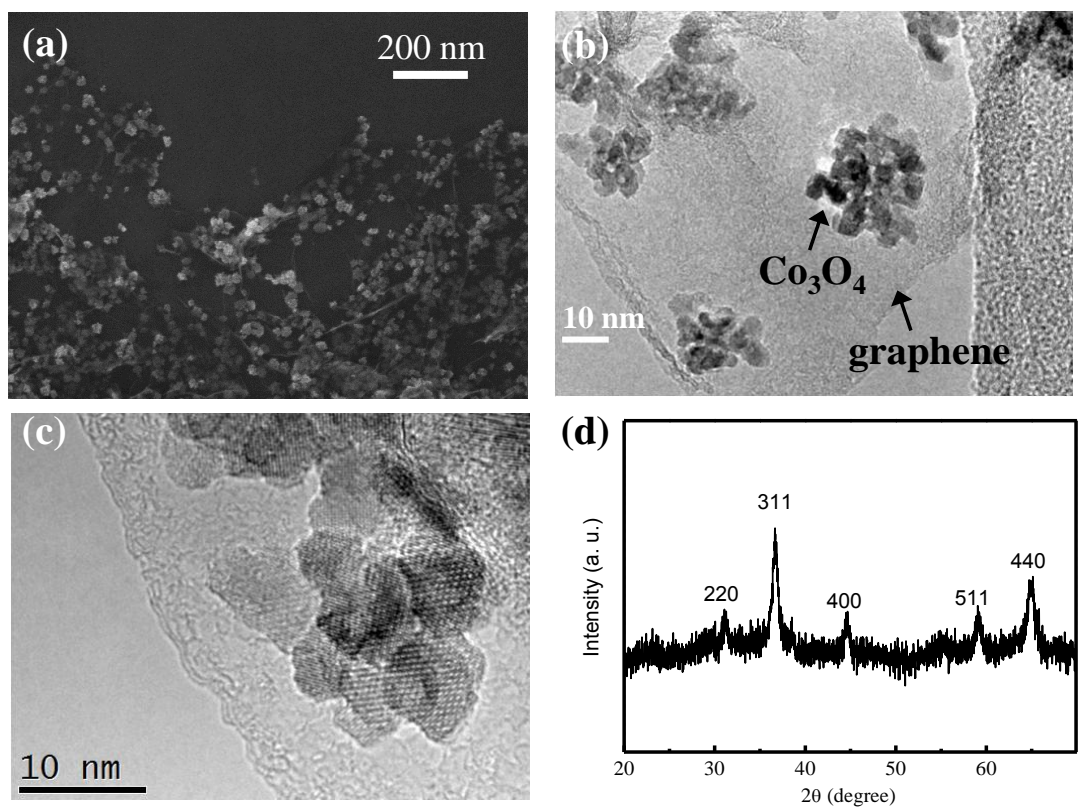

**Fig. S1.** $Co_3O_4$/rmGO hybrid prepared by the two step reaction. (a) SEM image of $Co_3O_4$/rmGO hybrid deposited on silicon substrate from a suspension in solution. (b) Low magnification and (c) high magnification TEM images of $Co_3O_4$/rmGO hybrid. (d) XRD spectrum of $Co_3O_4$/rmGO hybrid film. Transmission electron microscopy (TEM) revealed smaller particles in $Co_3O_4$/N-rmGO (~4-8 nm in size, Fig. 1b) than $Co_3O_4$/rmGO (~12-25 nm in size) shown here in (b) and (c).

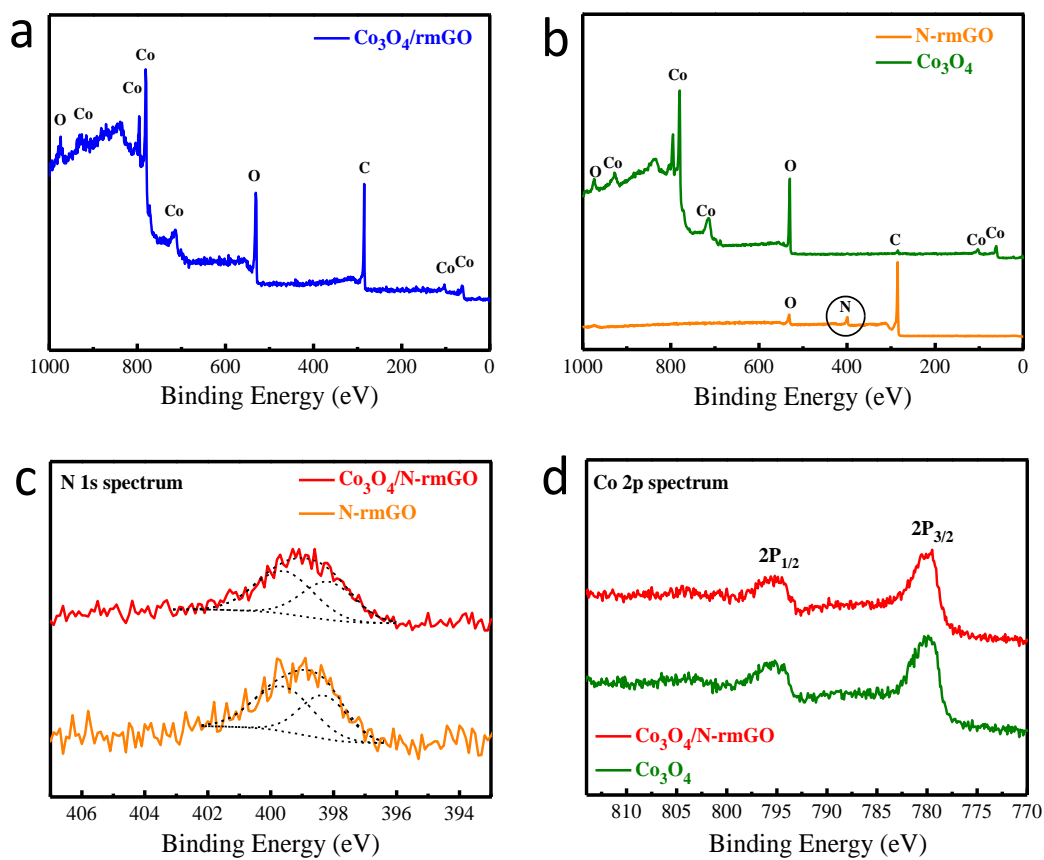

**Fig. S2.** XPS spectra of (a) $Co_3O_4$/rmGO hybrid and (b) $Co_3O_4$ nanocrystal and N-rmGO. N-rmGO was made through the same steps as making $Co_3O_4$/N-rmGO without adding any Co salt in the first step. This produced N-doped reduced GO with N clearly resolved in the GO sample by XPS. (c) High resolution N 1s spectra of $Co_3O_4$/N-rmGO hybrid and N-rmGO. (d) High resolution Co 2p spectra of $Co_3O_4$/N-rmGO hybrid and $Co_3O_4$ nanocrystal. The XPS spectra confirmed that N-dopants were on reduced GO sheets and not in $Co_3O_4$ nanocrystals. High resolution XPS spectra of the N peak revealed pyridinic and pyrrolic nitrogen species in $Co_3O_4$/N-rmGO and in N-rmGO shown here in (c).

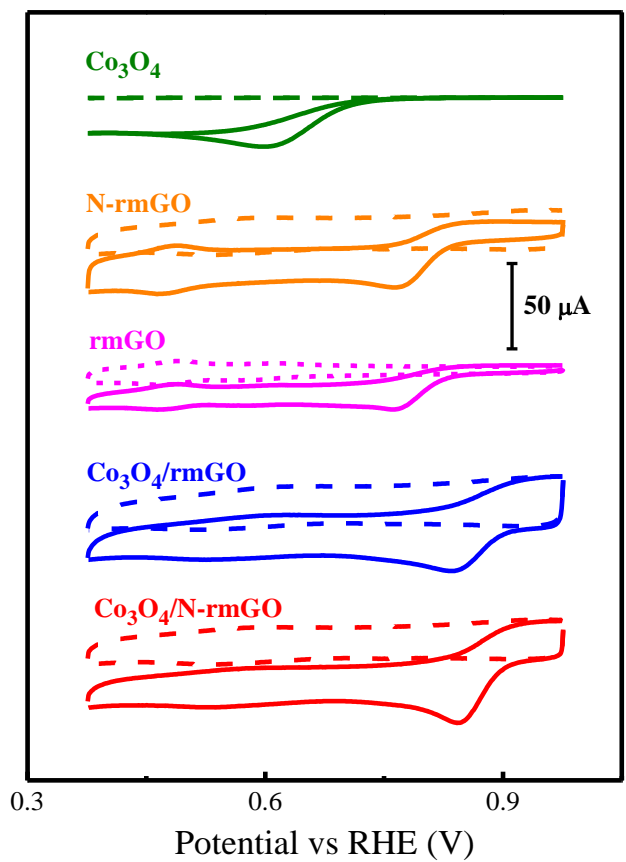

**Fig. S3.** CVs of Co₃O₄ nanocrystal, rmGO, N-rmGO, Co₃O₄/rmGO and Co₃O₄/N-rmGO (all loaded on glassy carbon electrodes with the same mass loading) in oxygen (solid) or argon (dash) saturated 0.1 M KOH. Free Co₃O₄ nanocrystals, rmGO or N-rmGO alone exhibited very poor ORR activity. In contrast, Co₃O₄/rmGO and Co₃O₄/N-rmGO hybrids showed much more positive ORR onset potentials and higher cathodic currents, suggesting synergistic ORR activity of the hybrid.

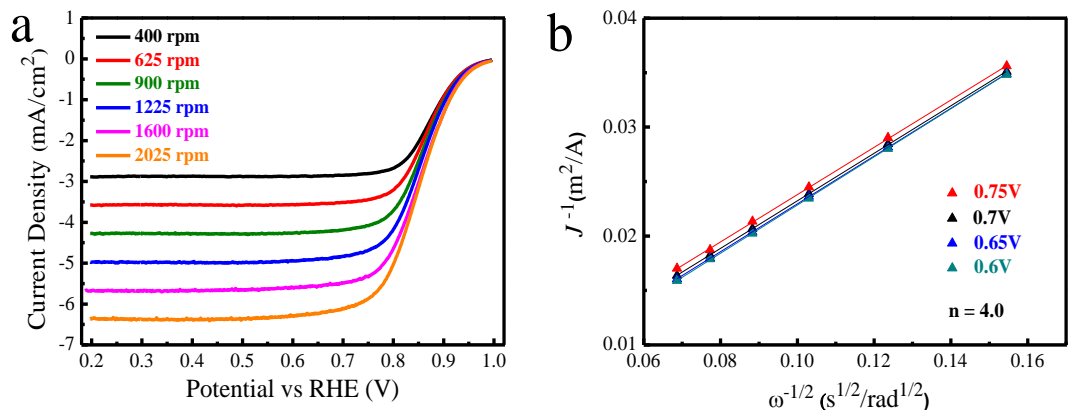

**Fig. S4.** . (a) Rotating-disk voltammogram of Pt/C in $O_2$-saturated 0.1 M KOH at a sweep rate of 5 mV/s and different rotation rates. The catalyst loading was 0.1 mg/cm². (b) Corresponding Koutecky–Levich plot ($J^{-1}$ versus $\omega^{-0.5}$) at different potentials.

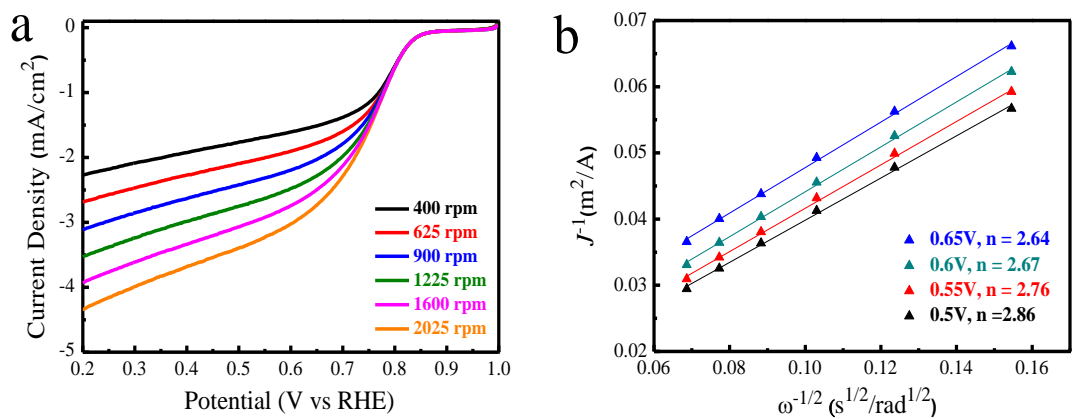

**Fig. S5.** (a) Rotating-disk voltammogram of N-rmGO (loading 0.1 mg/cm$^2$) in O$_2$-saturated 0.1 M KOH at a sweep rate of 5 mV/s and different rotation rates. (b) Corresponding Koutecky–Levich lot ($J^{-1}$ vs. $\omega^{-0.5}$) at different potentials.

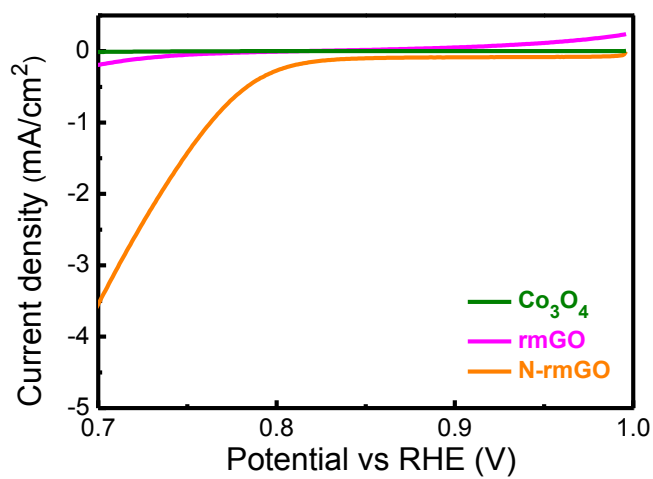

**Fig. S6.** Oxygen reduction currents of $Co_3O_4$ nanocrystal, rmGO and N-rmGO on carbon fiber paper in $O_2$-saturated 0.1 M KOH. The sample loadings were 0.24 mg/cm$^2$. At 0.7 V, $Co_3O_4$ nanocrystal, rmGO and N-rmGO afforded an ORR current density of 0.012, 0.19 and 3.5 mA/cm$^2$ respectively, which were 1-3 orders of magnitude lower than hybrids ($Co_3O_4$/rmGO - 12.3 mA/cm$^2$ and $Co_3O_4$/N-rmGO - 52.6 mA/cm$^2$ as shown in Fig. 4a), confirming a synergetic coupling between two catalytically non-active components in our hybrids for ORR catalysis.

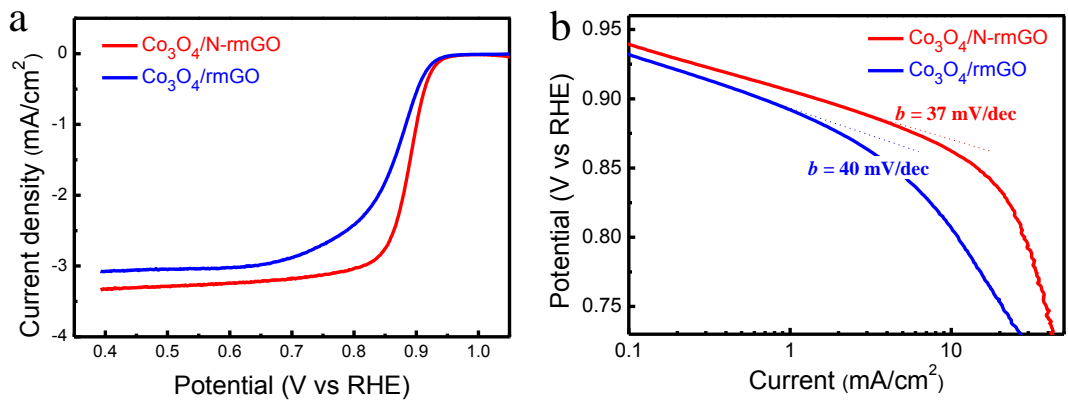

**Fig. S7.** (a) Rotating-disk voltammograms of $Co_3O_4$/rmGO hybrid and $Co_3O_4$/N-rmGO hybrid (loading 0.1 mg/cm$^2$) in $O_2$-saturated 1 M KOH with a sweep rate of 5 mV/s and rotation rate of 1600 rpm. (b) Tafel plots of $Co_3O_4$/rmGO and $Co_3O_4$/N-rmGO hybrids derived by the mass-transport correction of corresponding RDE data.

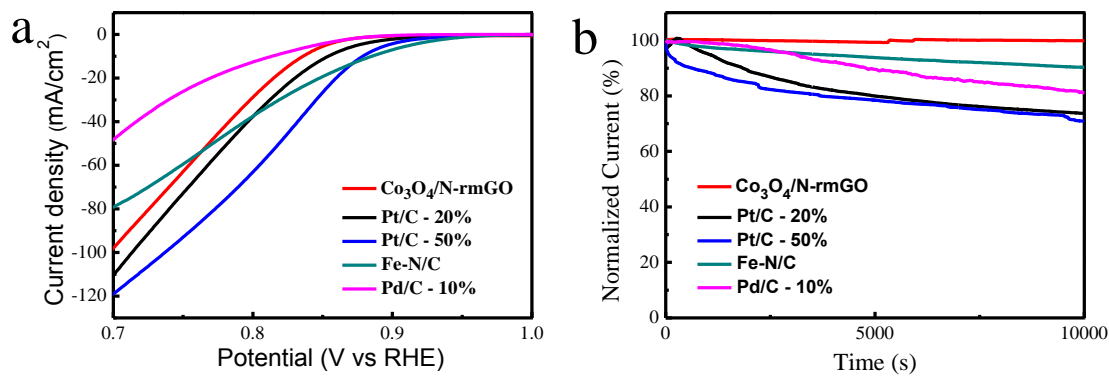

**Fig. S8.** ORR performance and stability of catalysts. (a) Oxygen reduction polarization curves of $Co_3O_4$/N-rmGO, Pt/C-20% (20 wt% Pt on Vulcan XC-72, Fuel Cell Store), Pt/C-50% (50 wt% Pt on Vulcan XC-72, Fuel Cell Store), Fe-N/C (Prepared followed Ref. [3] method) and Pd/C-10% (Palladium, 10% on activated carbon powder, Alfa Aesar) catalysts (catalyst loading ~0.24 mg/cm$^2$ for all samples) dispersed on carbon fiber paper (CFP) in $O_2$-saturated 1 M KOH electrolyte. (b) Chronoamperometric responses (percentage of current retained vs. operation time) of $Co_3O_4$/N-rmGO, Pt/C-20%, Pt/C-50%, Fe-N/C and Pd/C-10% on carbon fiber paper electrodes kept at 0.70 V vs. RHE in $O_2$-saturated 1M KOH electrolytes respectively. $Co_3O_4$/N-rmGO hybrid showed comparable ORR catalytic activity to Pt/C and superior stability in alkaline solutions.

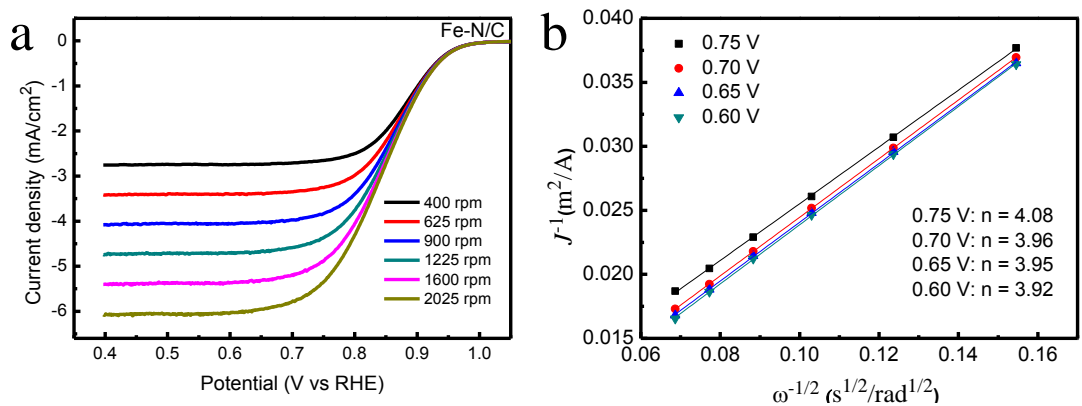

**Fig. S9.** (a) Rotating-disk voltammogram of Fe-N/C in $O_2$-saturated 0.1 M KOH at a sweep rate of 5 mV/s and different rotation rates. The catalyst loading is 100 μg/cm². (b) Corresponding Koutecky–Levich plot ($J^{-1}$ versus $\omega^{-0.5}$) at different potentials. The used Fe-N/C catalyst is high quality, matching the performance reported in the literature[3].

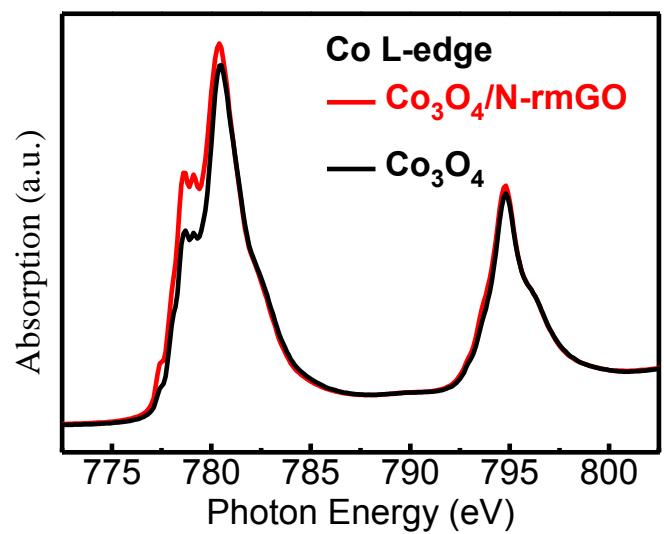

**Fig. S10.** Co L-edge XANES of $Co_3O_4$ nanocrystal and $Co_3O_4$/N-rmGO hybrid. The increase in the normalized peak area in hybrid compared to $Co_3O_4$ nanocrystal indicates increase of unoccupied Co 3d projected state in hybrid, suggesting lower electron density of Co site in hybrid.

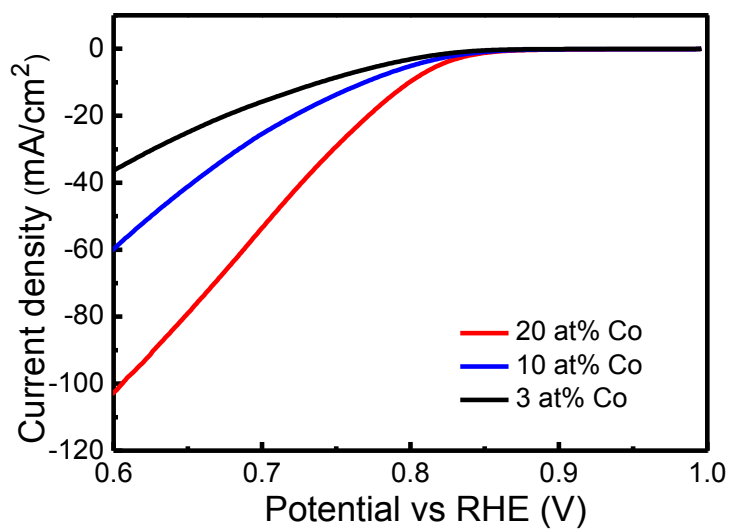

**Fig. S11.** Oxygen reduction currents of $Co_3O_4$/N-rmGO hybrids with various Co contents dispersed on carbon fiber paper in $O_2$-saturated 0.1 M KOH. The sample loading was 0.24 mg/cm$^2$. Lowering Co loading from 20 at% to 3-10 at% led to systematic reduction in ORR activity, suggesting that the active reaction sites in hybrid materials could be Co oxide species interfaced with GO.

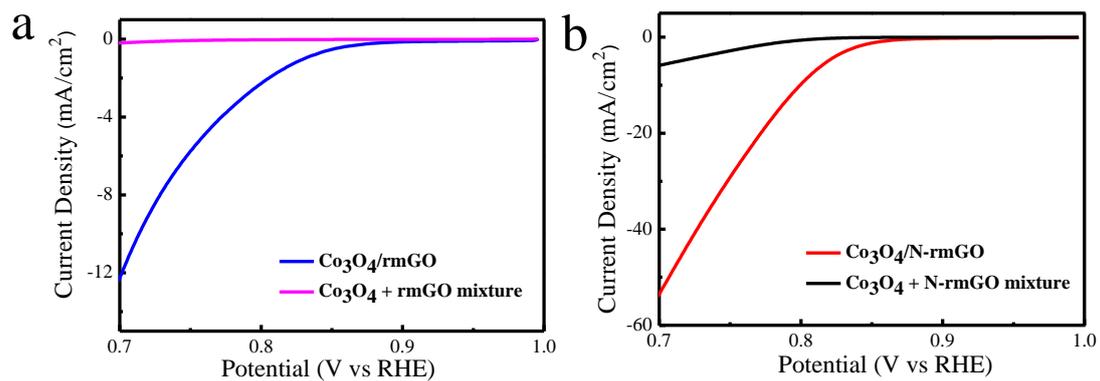

**Fig. S12.** Oxygen reduction currents of hybrids versus mixtures dispersed on carbon fiber paper in $O_2$-saturated 0.1 M KOH. The sample loading was 0.24 mg/cm$^2$. Both rmGO and N-rmGO hybrids showed much high activity compared to corresponding physical mixtures, confirming the importance of intimate interaction in the hybrid materials for ORR performance.

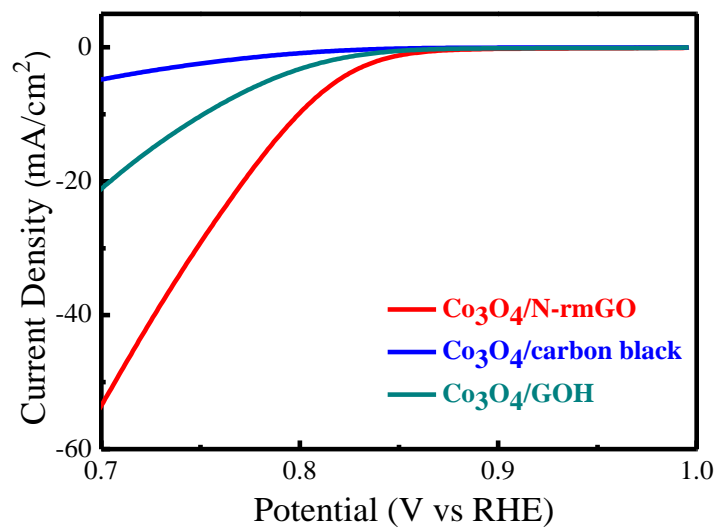

**Fig. S13.** ORR performance of $Co_3O_4$ hybrid catalysts with different carbon materials dispersed on carbon fiber paper in $O_2$-saturated 0.1 M KOH electrolyte (catalyst loading ~0.24 mg/cm$^2$ for all samples). The nitrogen doped hybrid prepared from Hummer's GO (GOH) showed lower activity than $Co_3O_4$/N-rmGO hybrid, which could be due to the lower conductivity of GOH. Carbon black hybrid (composite) also showed much lower activity, likely due to the lack of functional groups in carbon black as anchored sites of the nanoparticles. These results indicated the high conductivity and surface area, as well as suitable anchoring sites of mGO are important for the high activity of the synthesized hybrid materials.

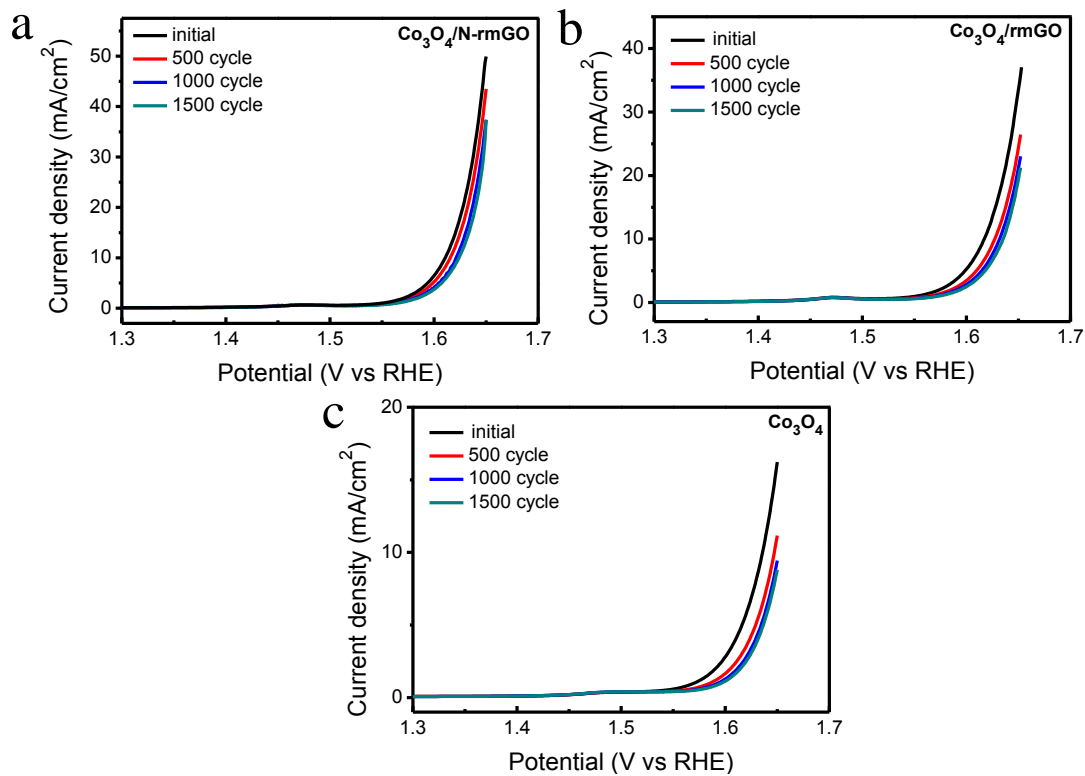

**Fig. S14.** OER stability test of $Co_3O_4$/N-rmGO, $Co_3O_4$/rmGO and $Co_3O_4$ catalysts dispersed on glassy carbon electrode (with 15% of Nafion as binder) in 1M KOH electrolyte (catalyst loading 0.10 mg/cm$^2$ for all samples). Cycles were swept between 1.25 V and 1.65 V at 0.2 V/s. The anodic sweeps showed in the figures were measured from 1.25 V to 1.65 V at 0.005 V/s with IR compensation. All three catalysts suffered certain current decrease in the beginning cycles (about 20-30% at 1.65 V), which is mainly due to the blockage of some active sites by the gradual accumulation of evolved $O_2$ bubbles. The OER current did not decrease significantly after 1000 cycles in all three catalysts, suggesting the $Co_3O_4$/N-rmGO and $Co_3O_4$/rmGO hybrid catalysts are inherently stable for OER.